\documentclass[apl,prb,showpacs,reprint,amsmath,amssymb,superscriptaddress,a4paper]{revtex4-1}
\usepackage[utf8]{inputenc}
\usepackage{amsmath}
\usepackage{amsfonts}
\usepackage{amssymb}
\usepackage{graphicx}
\usepackage{lineno}

\begin{document}
 \citestyle{unsrt}

\title{Scalable platform for nanocrystal-based quantum electronics}
\author{Joachim E. Sestoft$^{*}$}
\affiliation{Center for Quantum Devices \& Nano-science Center, Niels Bohr Institute, University of Copenhagen, 2100 Copenhagen, Denmark}
\thanks{These authors contributed equally\\ e-mail: joachim.sestoft@gmail.com, nygard@nbi.ku.dk}
\author{Aske N. Gejl$^{*}$}
\affiliation{NanoMade, Department of Physics, Technical University of Denmark, 2800 Kgs. Lyngby, Denmark}
\thanks{These authors contributed equally\\ e-mail: joachim.sestoft@gmail.com, nygard@nbi.ku.dk}
\author{Thomas Kanne$^{*}$}
\affiliation{Center for Quantum Devices \& Nano-science Center, Niels Bohr Institute, University of Copenhagen, 2100 Copenhagen, Denmark}
\thanks{These authors contributed equally\\ e-mail: joachim.sestoft@gmail.com, nygard@nbi.ku.dk}
\author{Rasmus D. Schlosser}
\affiliation{Center for Quantum Devices \& Nano-science Center, Niels Bohr Institute, University of Copenhagen, 2100 Copenhagen, Denmark}
\author{Daniel Ross}
\affiliation{Center for Quantum Devices \& Nano-science Center, Niels Bohr Institute, University of Copenhagen, 2100 Copenhagen, Denmark}
\author{Daniel Kj\ae r}
\affiliation{Center for Quantum Devices \& Nano-science Center, Niels Bohr Institute, University of Copenhagen, 2100 Copenhagen, Denmark}
\author{Kasper Grove-Rasmussen}
\affiliation{Center for Quantum Devices \& Nano-science Center, Niels Bohr Institute, University of Copenhagen, 2100 Copenhagen, Denmark}
\author{Jesper Nyg\aa rd}
\affiliation{Center for Quantum Devices \& Nano-science Center, Niels Bohr Institute, University of Copenhagen, 2100 Copenhagen, Denmark}

%%% Total manus length ~ 3495 words
\begin{abstract} 

%% New abstract 8 october 2021. Number of words: 155

Unlocking the full potential of nanocrystals in electronic devices requires scalable and deterministic manufacturing techniques. A platform offering promising alternative paths to scalable production is microtomy, the technique of cutting thin lamellae with large areas containing embedded nanostructures. This platform has so far not been used for fabrication of electronic quantum devices. Here, we combine microtomy with vapor-liquid-solid growth of III/V nanowires to create a scalable platform that can deterministically transfer large arrays of single and fused nanocrystals — offering single unit control and free choice of target substrate. We fabricate electronic devices on cross-sectioned InAs nanowires with good yield and demonstrate their ability to exhibit quantum phenomena such as conductance quantization, single electron charging, and wave interference. Finally, we devise how the platform can host rationally designed semiconductor/superconductor networks relevant for emerging quantum technologies.

\end{abstract}

\maketitle

Processing of nanocrystals (NCs) in scalable and deterministic ways is critical for their use in research fields ranging from new types of transistors\cite{xiang2006ge, nw_elec_opto_review_2006, yan2011programmable} to biosensors,\cite{cui2001nanowire, nw_elec_opto_review_2006, biosens_review2011} optoelectronics\cite{nw_elec_opto_review_2006, III-Vnw_optoreview_2011, nwlasers_Review_2016, nwsolarcell_review_2017} and modern quantum devices.\cite{prada2020andreev, giustino2021, flensberg2021engineered}

Generally, there are two approaches to scalable processing of NCs. One route is substrate dependent; here the NCs are either grown or etched from specific substrates, which also support later fabrication steps (e.g. circuit manufacturing). The second method is predominantly transfer-based; here the NCs are first synthesized and then later transferred to a specialized substrate. The techniques favored for the substrate specific approaches are vapour-liquid-solid growth (VLS), metalorganic vapour-phase-epitaxy, selective area epitaxy, Stranski-Krastanov growth and anisotropic etching. These techniques achieve excellent crystal quality and good control of position, but since devices are fabricated on the growth substrates, device performance can suffer from substrate-induced strain, substrate/NC dislocations, short-circuiting or sub-optimal conditions for heterostructure growth. Promising transfer-based approaches are Langmuir-Blodgett deposition,\cite{tao2008langmuir, collet2015large} nanocombed deposition,\cite{yao2013nanoscale} dry transfer printing\cite{mcalpine2007highly} and capillary force assembly\cite{flowaligningNWs_2001} which allow for high precision placement and good yield. However, they are often solution-based and depend on intensive preparation of the target substrates which influences device architectures and inhibits tracking of individual units. 

%Often, these techniques are also limited to certain types of materials since they may rely on NC alignment by external fields. 

An alternative route seeking to combine the best of both approaches is nanoskiving - the technique of embedding materials in a resin and cutting extremely thin samples by an ultramicrotome. Prior works have yielded several attractive and scalable nanostructures for use in optical and electrical applications.\cite{xu2008nanoskiving, lipomi2011use} As an example, fabrication of complex optically active nanostructures covering areas of several mm$^2$ has been demonstrated,\cite{xu2007fabrication} and also, nanoskiving was combined with etching of core-shell nanowires to form AlGaAs nanocylinders with tunable dimensions for optical devices.\cite{watson2014nanoskiving} Nevertheless, nanostructures produced by nanoskiving have not been implemented into quantum electronics. 

\begin{figure*}[tb!]
\includegraphics[scale=1]{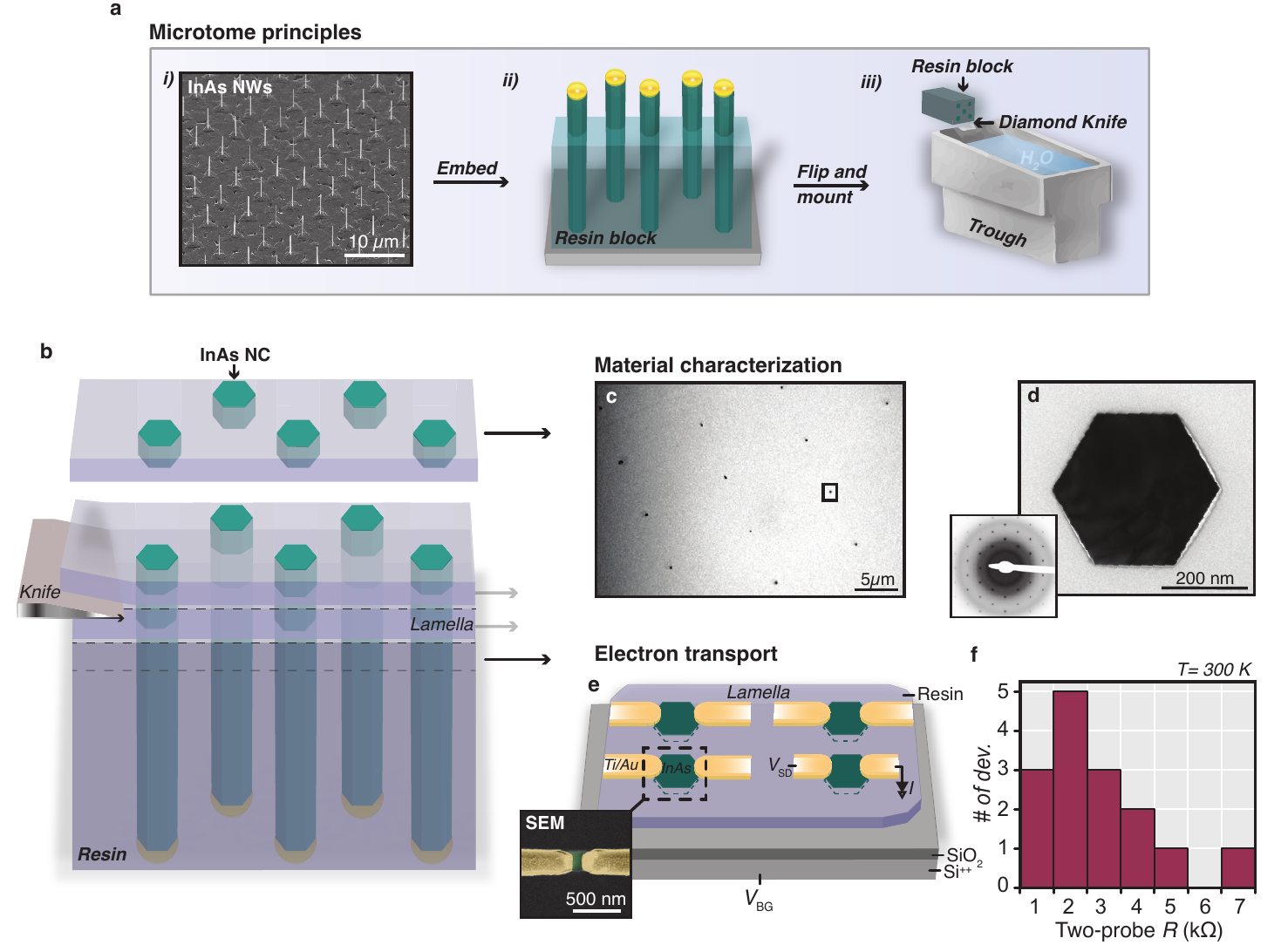}
\caption{\textbf{Scalable nanocrystal platform.} a) Illustrated microtome principles. $i$) SEM image of array of as-grown InAs nanowires. $ii$) Array of nanowires after embedding. $iii$) The resin block containing nanowires mounted next to the knife/trough. b) Zoom in on ($iii$) showing how subsequently cut lamellas can be transferred to different substrates (e.g. a Si substrate and a TEM grid). c) Low-magnification TEM micrograph of array of NCs transferred to a TEM grid. Black box indicates a single NC in the array. d) TEM image of a single NC. Inset: SAED pattern showing the intact crystallinity of the transferred NC. e) Schematic showing a lamella transferred to a Si/SiO$_2$ substrate, where four transferred NCs are electrically contacted. Inset: False-colored SEM image of a measured device. Green, InAs; Yellow, Ti/Au. f) Histogram of resistances ($R$) recorded on 15 conducting devices at $T =$ 300 K.} 
\end{figure*}

In this work we present a flexible platform for the transfer, and positioning of arrays of high quality NCs and NC networks to almost any substrate - achieved by combining microtomy with VLS growth of III/V nanowires. We measure electronic NC devices operated as field effect transistors (FETs) and find excellent mobilities as expected for InAs-based transistors, as well as quantum transport signatures when devices are operated at cryogenic temperatures. We show how the platform can be extended to host complex networks and semiconductor/superconductor devices with the prospect to realize many new types of gate-tunable superconducting qubits.\cite{prada2020andreev,aguado2020perspective, kjaergaard2020superconducting, giustino2021} In addition, our invention may allow for alternative routes in research fields such as deterministic placement of quantum dots for realization of the Hubbard model,\cite{hubbard1963electron} layering of optoelectronic structures/devices, foldable electronics, wireless single-electron logic controlled by electrical fields\cite{korotkov1995wireless} and single photon sources based on III/V quantum dots.\cite{mantynen2019single}

\section*{Scalable nanocrystal platform}

%We point out that other types of microtomes do not rely on liquid-based transferring, but instead use solid stages providing high control of positioning.\cite{}

In Figure 1 we show the principles of the scalable NC device platform. InAs nanowires are grown on (111)B InAs substrates by the VLS process where the position of the nanowires are controlled by placing Au catalyst particles via electron beam lithography (EBL). Following growth the substrate is cleaved into pieces that are compatible with the microtome setup and embedded in a resin. The resin block containing the positioned nanowires is detached from the growth substrate and is mounted and aligned in the microtome setup. This work flow is illustrated in Figure 1a. Using the microtome setup we cut thin ($\sim$80-100 nm) slices of resin (lamellas), containing positioned NCs as seen schematically in Figure 1b. The lamellas can be transferred to almost any type of substrate via a liquid medium kept in the trough (knifeboat). Figure 1c shows a bright-field (BF) transmission electron microscope (TEM) image of an array of NCs with the positioning translated from the growth substrate to the TEM grid. A close-up TEM micrograph of an individual NC with a clear hexagonal morphology is seen in Figure 1d. The inset shows the selected area electron diffraction (SAED) pattern where the regularly spaced spots indicate that the crystal remains intact after the sectioning/transfer process, and the grey rings correspond to the amorphous carbon grid. Additional information on the use of a ultramicrotome to create lamellas with arrays of NCs can be found in the Supporting Information, Section 1. 

% and its corresponding TEM image showing a NC from the previously cut lamella (scale bar is 200 nm)

\begin{figure*}[htb!]
\includegraphics[scale=1]{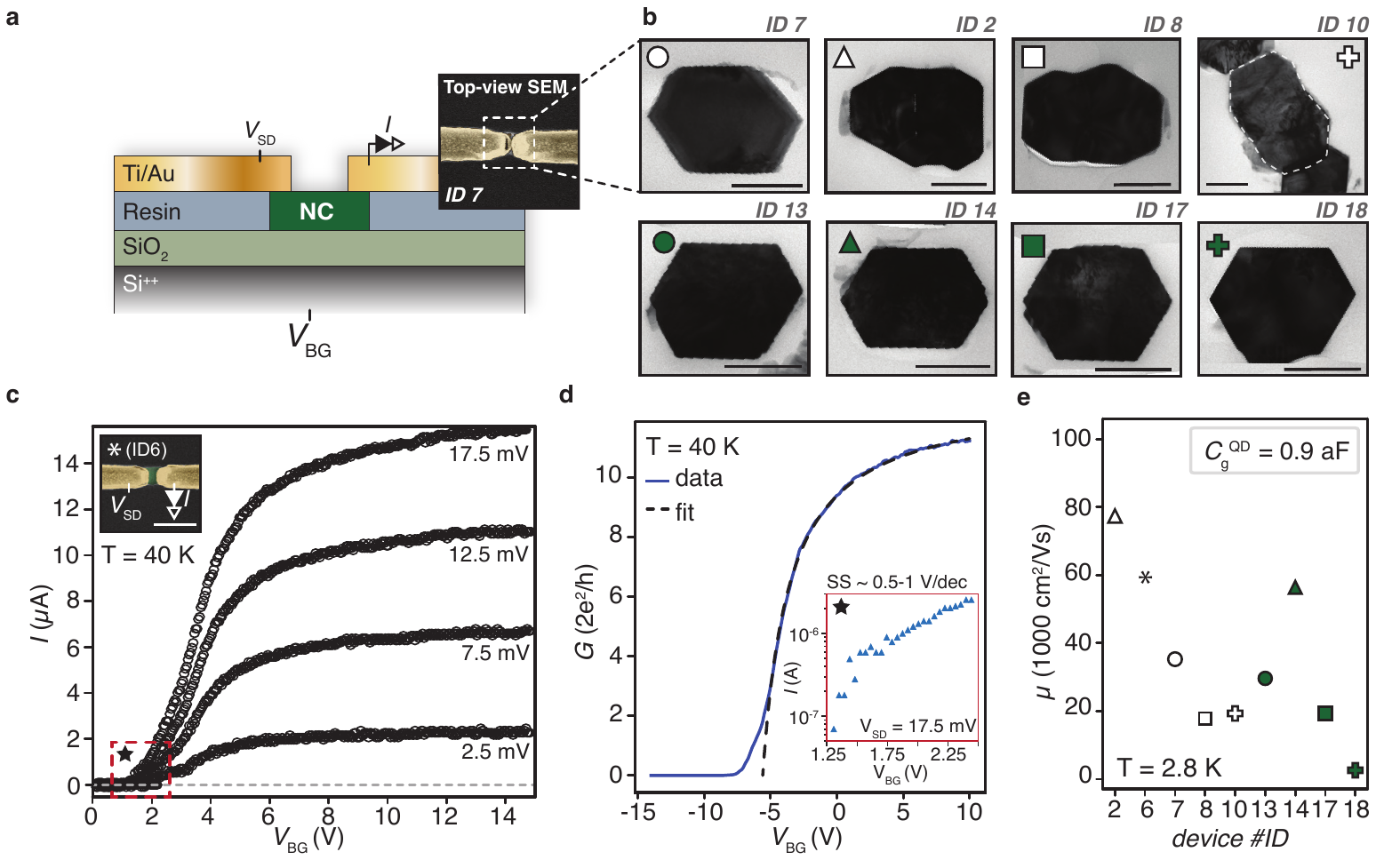}
\caption{\textbf{Nanocrystal FET behavior.} a) Side-view illustration of a NC FET device. Inset: False-colored SEM image of a NC device. Yellow, Ti/Au contacts; green, NC; $V_{\mathrm{SD}}$, the applied voltage bias; $I$, measured current; $V_{\mathrm{BG}}$, backgate voltage shifting the chemical potential of the NC. b) TEM images of corresponding NCs from a lamella cut before the lamella used for fabrication of FETs. All scalebars are 200 nm. c) Current as a function of $V_{\mathrm{BG}}$ recorded at $V_{\mathrm{SD}}$ = 2.5, 7.5, 12.5, 17.5 mV. Corresponding device has ID 6 and is shown in the inset. d) FET model (dashed line) fitted to the linear conductance ($G$) swept as a function of $V_{\mathrm{BG}}$ (blue). Inset: $I$ plotted vs $V_{\mathrm{BG}}$ on a semi-logarithmic scale. Trace marked by a star in (c). d) Estimated FE mobilities for all gateable devices.} 
\end{figure*}

Following the initial transfer to a TEM grid, we transfer the subsequently cut lamella to a Si$^{++}$/SiO$_{\mathrm{2}}$ substrate. The NCs in the lamella are contacted electrically using standard cleanroom fabrication techniques as seen schematically in Figure 1e, where an array of four NCs are contacted. Details on fabrication can be found in the Methods section. The NCs are kept in plane with the lamella and the Ti/Au contacts are placed on top of the resin, resulting in the NCs being electrically accessed from the top cut facet. Since the resin is insulating the NCs are electrically isolated from each other and one can apply a voltage bias, $V_{\mathrm{SD}}$, across the individual NCs and measure the current, $I$. The inset in Figure 1e shows a false-colored scanning electron microscope (SEM) image of one of the measured devices. In Figure 1f we show a histogram of the two-probe resistances of 15 working devices out of 17 measured at room temperature. The resistances are mostly similar in the few k$\Omega$-regime showing that ohmic contacts can be reproducibly fabricated.

% A key advantage of this platform, is its unique ability to correlate electrical measurements to the NC material quality while keeping the scalable nature of the platform intact. Under optimized growth parameters nanowires are uniform in composition and morphology along the growth direction, which results in NCs cut from the same nanowire to be near-indistinguishable from each other. Since they are grown in arrays, a NC which has been electrically contacted can readily be compared to the previously or later cut NC and be investigated by e.g. SEM, TEM or atomic force microscopy (AFM). At current fabrication stages, the top/bottom facets of consecutively cut NCs may recurrently appear slightly irregular. We believe this effect to be effectively mitigated by optimization of the alignment of the nanowire crystal axes with the cutting plane of the knife ensuring accurate crystal cleaving. Further data and analysis of the cut face quality can be seen in the Supporting Information, Section 2. Assuming optimized growth and fabrication procedures, we expect that this approach can be utilized for quality control of arrays of quantum devices compatible with large-scale manufacturing processes. 

Concluding on this section, we expect the scalable NC platform to easily be extended to a large number of materials relevant for fabrication of quantum devices, as materials such as Si, GaAs, BiTe, Al, AlO$_2$, Cu and Au have already been investigated by commercial ultramicrotome manufacturers,\cite{leicaimgs2021} and highly relevant materials such as Al, Pb, Sn, As, In, Ga, Ge, Sb, Bi, SiO$_2$ among many others are either proven or predicted to be intact after microtoming.\cite{lipomi2011use}

%\newpage

\section*{Nanocrystal FET\lowercase{s} }

Having demonstrated the scalable device platform we explore the field effect behavior in a series of individual and merged InAs NCs. In Figure 2a we show a cross-section schematic of a single NC FET device with an inset showing a false-colored SEM image of one of the measured devices. At cryogenic temperatures 9 of the 15 devices were still conducting and exhibited a field effect response. In Figure 2b we show the TEM images of the NCs corresponding to the device-NCs of which transport data is presented. As an example, we show the SEM image of a device fabricated on a NC (ID7) and a TEM image of a NC from the same nanowire. We investigate both NCs sectioned from single nanowires (ID6, 7, 13, 15, 17, 18), as well as devices that were fabricated on two or more merged NCs (ID2, 8, 10). One NC (ID6) is not shown due to unfortunate lamella placement on the TEM grid. The merged NCs were cut from nanowires that were deterministically grown together during a radial growth step as described in the final section. Details can be found in the Methods section and recently published work.\cite{kanne2021double} The hexagonal shapes of the single NCs appear somewhat irregular, which is an artifact from the close spacing of the single nanowires grown under unoptimized conditions. As inter-nanowire spacings decrease local growth parameters change due to temperature fluctuations and variations in shared collection areas. 

In Figure 2c we show current, $I$, measured as a function of backgate voltage, $V_{\mathrm{BG}}$, at different bias voltages, $V_{\mathrm{SD}}$, of a NC device (ID6) at low temperature, $T$~=~40~K. We choose this temperature since most conductance resonances are averaged out by thermal effects as shown in Supporting Information, Section 3. By sweeping $V_{\mathrm{BG}}$ we modulate the number of carriers in the semiconductor and change the current flowing across the NC, bringing the device through three distinct regions. In the range $V_{\mathrm{BG}}$= 0--2 V no current is running corresponding to an OFF state. From $V_{\mathrm{BG}}$ = 2--4 V the current increases linearly, while the current saturates above $V_{\mathrm{BG}}$ = 4 V. The saturation current increases linearly with $V_{\mathrm{SD}}$ implying that high quality ohmic contacts to the NCs have been fabricated, as expected from n-type InAs-based electronic devices.\cite{thelander2004electron} 

\begin{figure*}[htb!]
\includegraphics[scale=1]{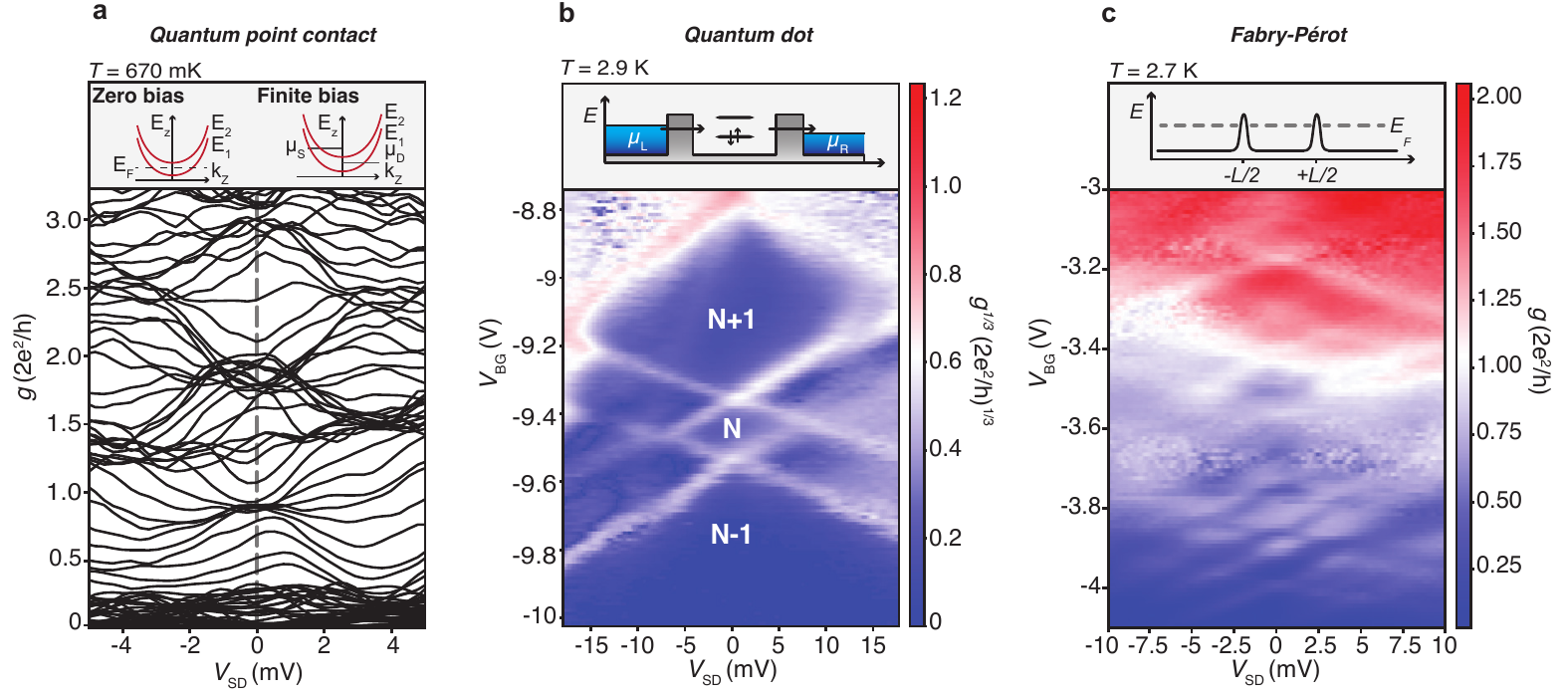}
\caption{\textbf{Quantum behavior in nanocrystal devices.} a) Zero-field waterfall plot of differential conductance, $g = dI/dV_{\mathrm{SD}}$, as a function of $V_{\mathrm{SD}}$ and $V_{\mathrm{BG}}$ showing an approximate bunching of curves at (half) integer values of 2$\mathrm{e}^2/\mathrm{h}$ at $V_{\mathrm{SD}}$ = 0 ($|V_{\mathrm{SD}}|>$ 2 mV). Inset: Zero-bias and finite bias band diagram of a one-dimensional quantum point contact. b) Color-plot of $g$ versus $V_{\mathrm{SD}}$ and $V_{\mathrm{BG}}$ showing a series of Coulomb diamonds. The conductance is plotted in the third root to emphasize low values. Inset: Energy diagram of the quantum dot charging process. c) Colour-plot of $g$ as a function of $V_{\mathrm{SD}}$ and $V_{\mathrm{BG}}$ showing Fabry-Pérot oscillations. Inset: Energy diagram with opaque barriers and position of Fermi level, $E_\textrm{F}$.}
\end{figure*}

We now shift our focus towards the mobility of the NCs. The FET mobility is estimated using a commonly used expression for nanowire FETs\cite{G_l_2015}. Details on fitting are found in Supporting Information, Section 5. We apply this model to underline the overall FET behavior of NC devices. Figure 2d shows the linear conductance measured ($G$) as function of $V_{\mathrm{BG}}$ where the corresponding fit follows the behavior of the FET excellently until close to pinch-off. The inset shows a zoom-in on the subthreshold swing, SS, of the $I/V_{\mathrm{BG}}$ curve at $V_{\mathrm{SD}}$ = 17.5 mV from Figure 2c. Depending on the chosen range, the SS varies between $\sim$0.5-1 V/dec which is significantly lower than state-of-the-art nanowire FETs.\cite{salahuddin2008use, bryllert2005vertical} This behavior may be explained by the extremely short distance between the contacts resulting in short channel effects which are known to reduce the SS in ultra short devices.\cite{fiegna1994scaling} Additional hysteresis plots are shown in the Supporting Information, Section 4, and are comparable to other types of nanowire-based FETs fabricated on Si$^{++}$/SiO$_{2}$ substrates.\cite{G_l_2015}

% $$G(V_{\mathrm{BG}})=\bigg(R_{\mathrm{s}}+\frac{W^2}{\mu \cdot C_{\mathrm{G}} \ (V_{\mathrm{BG}}-V_{\mathrm{th}})}\bigg)^{-1}$$, here $G(V_{\mathrm{BG}})$ is the linear conductance, $R_{\mathrm{s}}$ is the series resistance, $W$ is the distance between the contacts, $\mu$ is the FET mobility, $C_{\mathrm{G}}$ is the gate/NC capacitance and $V_{\mathrm{th}}$ is the threshold voltage. We use $\mu$, $R_{\mathrm{s}}$ and V$_{\mathrm{th}}$ as the free fit parameters, and estimate $C_{\mathrm{G}}$ from the slopes of the Coulomb diamond shown in Figure 3b (this value is used across all measured devices) and measure the distance between the contacts, $W$, by SEM. 

We have estimated the low temperature mobilities for all 9 NC FETs as seen in Figure 2e ($T = 2.8$ K). Measurements used to obtain the mobilities and their according fits are shown in the Supporting Information, Section 5. We observe no noticeable trend between the estimated mobilities obtained from NCs cut from a single nanowire or merged nanowires indicative of no significant barrier in the merged NCs. This may be explained by specific growth conditions causing the merged NCs to effectively act as a single crystal, hence suggesting that formation of larger networks based on current growth parameters is feasible. The estimated mobilities seem comparable to or larger than values found in SAG and VLS grown InAs and InSb nanowires,\cite{G_l_2015, krizek2018field} although they vary quite substantially across the different devices. This may be a side-effect of large variances in capacitance, and the discrepancy between the model accounting for diffusive transport and some device lengths being comparable to or smaller than the experimentally estimated mean free path of InAs nanowires ($\sim$150 nm).\cite{ball_inas_2013} Hence, we use the FET model mainly to highlight the predictable field effect behavior of the NC devices. In summary, this section shows that the NC transfer platform can host ultra-small FET devices and that these behave predictably in electrostatic fields, which is a fundamental requirement for control of complex electronic quantum devices.

% , and only tentatively to compare with literature values of other nanowire-based FETs.
% We speculate that the subthreshold swing may be improved by looking into different types of dielectric resins, removal of the resin and/or encapsulations of the nanowires by a high-K dielectric before the embedding process. 
%If desirable, the short channel effects may be almost entirely avoided by merging numerous nanowires during overgrowth before embedding, effectively creating longer NCs as discussed in Figure 4.

%In summary, this section shows that the NC transfer platform can host ultra-small FET devices and that these behave predictably in electrostatic fields, which is a fundamental requirement for control of complex electronic quantum devices.

\section*{Quantum transport in single nanocrystal\lowercase{s}}

A prerequisite for using the NC platform for quantum-based applications is the capability to hold quantized phenomena in single NCs. Here we show a series of quantum effects that emerge around liquid He temperatures in a set of devices based on single NCs.

We plot differential conductance, $g = dI/dV_{\mathrm{SD}}$, as function of $V_{\mathrm{SD}}$ and $V_{\mathrm{BG}}$ in a waterfall plot for device ID7 as seen in Figure 3a. As we sweep $V_{\mathrm{BG}}$ and $V_{\mathrm{SD}}$ we observe a bunching of the individual traces close to the first few units of twice the conductance quantum ($2G_0 = 2\mathrm{e^2/h}$) around zero bias. At larger bias voltages ($|V_{\mathrm{SD}}|~\sim$~3--4 mV) this bunching effect is no longer observed at integer values but instead at half-integer values. These effects are discussed based on the inset shown in Figure 3a where a schematic of the parabolic dispersion relation of the first few sub-bands in an ideal one-dimensional quantum point contact (QPC) is seen. This textbook example assumes that the device length is much smaller than the mean free path ($l_{\mathrm{D}} \ll l_{\mathrm{mfp}}$), that the temperature of the system is smaller than the subband spacing ($k_{\mathrm{B}}T \ll E_2-E_1$) and that the Fermi wavelength is comparable to the width of the constriction ($\lambda_{\mathrm{F}} \sim W$). Changing the Fermi level in the case with zero bias ($\mu_{\mathrm{L}} = \mu_{\mathrm{R}}$) will modulate the energetically accessible modes supported by the QPC in an integer fashion where right and left moving carriers are kept in equilibrium thus increasing conductance in steps of $2G_0$. Conversely, by applying an offset in the bias ($\mu_{\mathrm{L}} \neq \mu_{\mathrm{R}}$) another mode can be accessed now exclusively by e.g. left moving carriers resulting in an increase in conductance by only one $G_0$. 

The presented data are in fairly good agreement with the described picture. We find that the distance between the contacts (determined by SEM to $W<$ 100 nm) is smaller than the typically experimentally determined mean free path in InAs nanowires ($l_{\mathrm{mfp}}^{\mathrm{InAs}} \sim 150$ nm\cite{ball_inas_2013}) and that the thermal energy of the system ($k_{B}T \sim 0.06$ meV) is much smaller than the estimated zero-field sub-band spacing ($E_{2}-E_{1}$) of $\sim$3-5 meV. However, as the width of the NC-device is larger than the Fermi wavelength typically found in InAs nanowires ($\lambda_{\mathrm{F}}^{\textrm{InAs}}\sim$ 20-30 nm\cite{jespersen2009mesoscopic}), the system is likely only quasi one-dimensional. We note that as we sweep $V_{\mathrm{BG}}$ the confinement potential of the electron path through the device may be changed. Hence by operating $V_{\mathrm{BG}}$ at negative voltages and taking into account screening from the leads a saddle-shaped confinement potential with a width that is comparable to $\lambda_{\mathrm{F}}^{\textrm{InAs}}$ may be formed in the device. This behavior and the quality of data is comparable to measurements on quantum point contact-like devices based on InAs nanowires.\cite{thelanderqpcqdot2004,ball_inas_2013, heedt2016inasqpc, heedt2017signatures} A total of three devices show zero-bias traces with according behavior as seen in the Supporting Information, Section 6.  

%This alternating magnitude of $E_{\mathrm{add}}$ is observed as the sizes of the diamonds change for the filling of the states.

% \newpage

When operating at larger negative $V_{\mathrm{BG}}$ the device is brought into the tunneling regime, likely due to opposing fields from the contacts acting as barriers for low carrier densities.\cite{kretinin2010fabry} In Figure 3b we show a color plot of $g$ versus $V_{\mathrm{SD}}$ and $V_{\mathrm{BG}}$ obtained from device ID7 at $T$ = 2.9 K. Here we observe diamond shaped conductance resonances with the dark blue regions inside the diamonds corresponding to transport being suppressed by Coulomb blockade. In the inset in Figure 3b we show an illustration of an energy diagram for a quantum dot (QD) tunnel-coupled to a source and a drain contact. When the chemical potential is modulated in the zero-bias case ($\mu_{\mathrm{L}} = \mu_{\mathrm{R}}$, not shown in illustration) single quantum mechanical states in the QD are accessed as the charging energy ($E_{\mathrm{C}}$) determined by the electrostatic force of the system is overcome. Applying a bias offset ($\mu_{\mathrm{L}} \neq \mu_{\mathrm{R}}$) to this model allows us to measure the addition energy ($E_{\mathrm{add}}$) as the source and drain potentials are aligned with the energy of the quantum state where the conductance resonances are observed. From the shape of the diamonds we extract the charging energy, $E_{\mathrm{C}}$ = 7.5 meV, total capacitance, $C_{\mathrm{tot}}$ = 22 aF, gate capacitance, $C_{\mathrm{g}}$ = 0.9 aF and lever arm $\alpha_{\mathrm{g}}$ = 0.03. The quantum confinement energy is tentatively estimated to $\Delta$ $\sim$ 12.5 meV. These findings are similar to values found in literature for a variety of self-assembled and nanowire-based InAs quantum dots.\cite{thelanderqpcqdot2004, jung2005qdot1, kretinin2010fabry, petta2010inasnwqdot, iju2017crystalqdot} Quantum dot like structures were also observed in device ID 6. Additional data on excited states are shown in Supporting Information, Section 7. 

% As the system dimensions approach the Bohr radius the quantum confinement energy ($\Delta$) associated with the filling of next state grows to the order of $E_{\mathrm{C}}$ and also needs to be considered. In a simplified picture the charging and discharging of a QD can be described by the addition energy, $E_{\mathrm{add}} = E_{\mathrm{C}} + \Delta$, where $\Delta$ is accounted for only for odd fillings due to electron spin degeneracy. 

\begin{figure*}[htb!]
\includegraphics[scale=1]{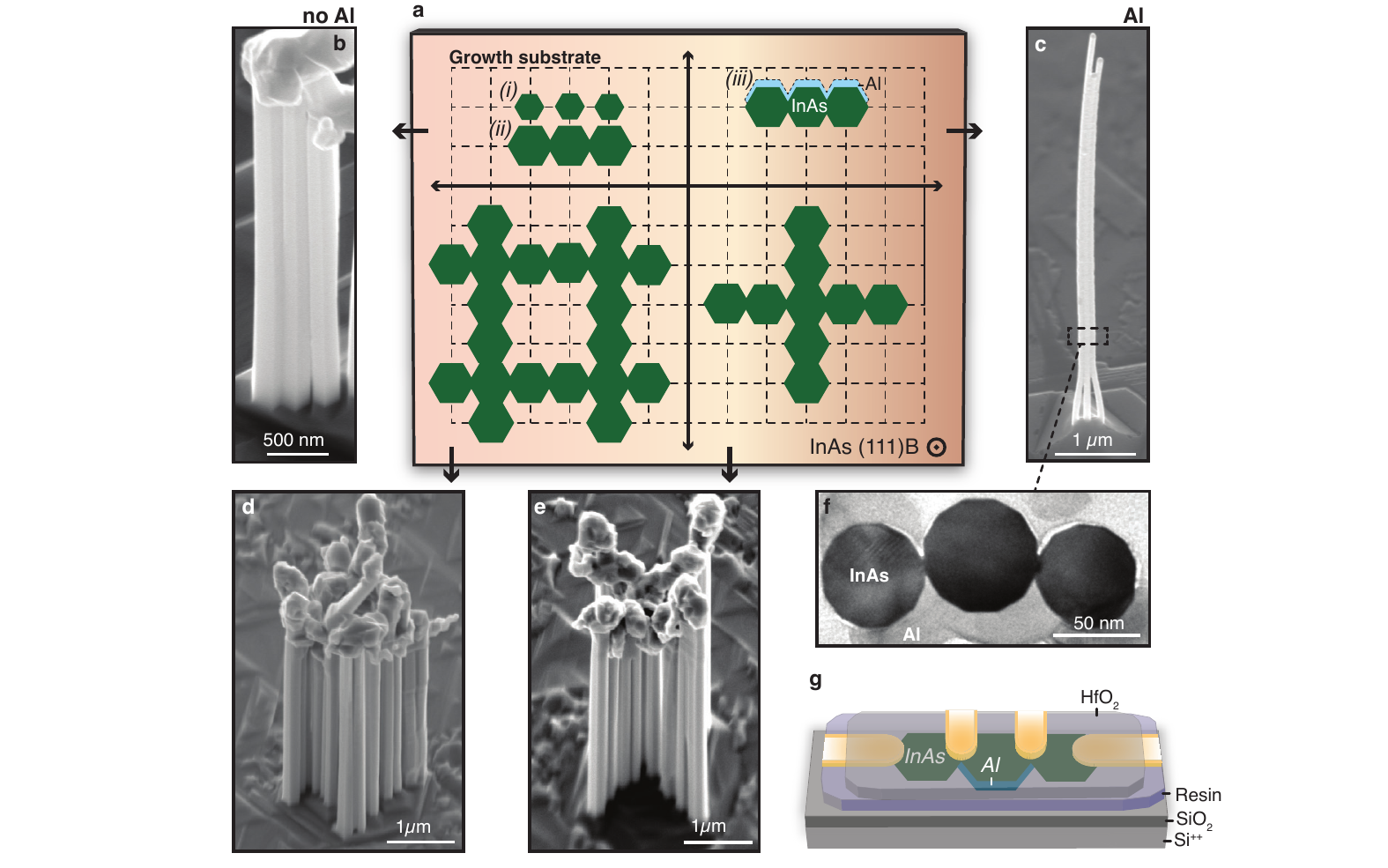}
\caption{\textbf{Nanocrystal networks and superconductors.} a) Schematic of an imagined growth substrate coordinate system. Top-view placement of the green hexagons illustrate where InAs nanowires can be (over)grown in order to form complex NC networks. b,c) SEM images of three merged nanowires without and with epitaxially grown Al, respectively. d) SEM image showing a series of merged nanowires forming a loop-structure. e) SEM image of merged nanowires that form a cross. f) Cross-section view TEM image of three merged NCs with epitaxially grown Al. g) Illustration of a hybrid InAs/Al NC device suited for superconducting quantum electronics.}
\end{figure*}

In Figure 3c we show conductance spectroscopy of device ID6 where a checker-board pattern on a background of conductance is seen. When electronic waves are backscattered from a set of opaque barriers (as seen in the inset in Figure 3c), likely residing at the NC/contact interface, resonant conductance features may appear as either the chemical potential or the potential landscape of the device is changed. This is analogous to the transmission/reflection of light waves in Fabry-Pérot cavities. By converting $C_{\textrm{G}}$ extracted from Figure 3b into capacitance per unit length ($C_{\textrm{G}}^{L} = 6$ pF/m), and taking the spacing between the oscillations at zero bias, $\Delta V_{\textrm{BG}}^{FP} =$ 250 mV, we can estimate the length of the Fabry-Pérot system to be $L = \frac{2e}{C_{\textrm{G}}^{L}\Delta V_{\textrm{BG}}^{FP}} \sim$ 215 nm,\cite{biercuk2005anomalous} which is comparable to the width of the NC measured by SEM. As we sweep $V_{\textrm{BG}}$ the finite conductance background is modulated and a large variation in $\Delta V_{\textrm{BG}}^{FP}$ is observed suggesting that the number of Fabry-Pérot modes present in the NC device changes. Similar behavior was observed in device ID2. These findings are comparable to electronic Fabry-Pérot interference found in nanowires\cite{duan2003single, kretinin2010fabry, heedt2016inasqpc} and nanotubes.\cite{liang2001fabry, jorgensen2006electron}

Concluding on Figure 3, we demonstrate the ability of the NC devices to hold three fundamental types of electronic quantum phenomena. 

% Taking into account that charge accumulates at the surface of InAs-based systems due to Fermi level pinning at vacuum,\cite{mead1963fermi, olsson1996charge} other material systems (such as InSb\cite{mead1963fermi}) which are less sensitive to surface disorder are likely to produce even more stable transport features. 

\section*{Networks and hybrid devices}

Recently, controlled growth of merged superstructures of nanowires has been demonstrated,\cite{kanne2021double} as well as semiconductor/superconductor nanowire heterostructures composed of III/V materials and Sn, Pb, and Al, known as hybrid devices.\cite{chang2015hard, gazibegovic2017epitaxy, sestoft2018engineering, carrad2020shadow, khan2020highly, pendharkar2021parity, kanne2021epitaxial} These emerging materials are in high demand as they are proposed to serve as a foundation for emerging quantum information technologies.\cite{prada2020andreev, giustino2021, flensberg2021engineered} Here we show how to extend the presented NC platform to these new material combinations and complex device architectures. 

% Keeping in mind that electron transport across merged NCs can readily be achieved, we move on to describe how to merge multiple nanowires along their growth direction and apply in-situ metal depositions. 

In Figure 4a a schematic of a growth substrate is presented where the dashed lines indicate a coordinate system. By use of standard EBL processes, catalyst particles can be placed on the growth substrate in predetermined patterns, and when VLS growth is initiated nanowires are formed at the locations of the catalyst particles.\cite{kanne2021double} Given the conditions under standard VLS growth (see Methods for details) the nanowires grow predominantly in the vertical direction with respect to the InAs (111)B substrate, and only minimally in their radial directions $\{1\bar{1}00\}$. Controlling the design of the catalyst particle placement, this can result in extremely closely spaced single nanowires, as illustrated in the cross-section view in Figure 4a ($i$). From arrays of single nanowires, one can merge single nanowires along their growth directions in order to form networks (see Figure 4a ($ii$) for cross-section schematic) in two ways. 

In the first way, single nanowires are grown at close distances from each other ($\sim 50-160$ nm) and from seed particles with large volumes. As the seed particle volume determines the width of the nanowires, this ensures that the nanowires are rigid. After reinitiating growth in a second growth step (see Methods) the nanowires grow predominantly in the radial directions. Depending on the placement of the single nanowires we can merge two or more nanowires into networks, and control the extent of which individual nanowires are merged. In Figure 4a ($ii$) and Figure 4b, we show a cross-section schematic and a SEM image of an example of three nanowires merged during radial overgrowth, respectively.

In the second way, nanowires are also grown closely spaced apart ($\sim 160$ nm) from seed particles with smaller volumes, such that the nanowires become thin. Local vibrations and van der Waals forces then make the nanowires touch and stick during the first growth phase, and allow the small amount of radial growth still present during the VLS mechanism to merge the individual nanowires. We provide examples of this in the SEM and cross-section TEM image shown in Figure 4c and d. After growth, these structures are embedded and sectioned as already described, creating networks of connected or disconnected semiconductor islands. Additionally, they can be grown with a superconductor (Al) as seen in the illustration in Figure 4a ($iii$) and the TEM image in Figure 4f. 

Based on the first scheme, we have designed and grown complex nanowire network structures as seen in Figure 4d and e. We highlight these particular structures as combinations of these after sectioning can form myriads of complex NC-based device structures, simply by controlling the placement of the seed particles. The erratic growth on the nanowire ends can be alleviated by optimization of growth parameters. 
In Figure 4g we showcase how the presented platform can be used to built a type of device (superconducting hybrid island) that has obtained widespread attention lately for research into Majorana physics in semiconductor/superconductor nanostructures.\cite{prada2020andreev, giustino2021, flensberg2021engineered} 

% Here we show SEM images of networks that after sectioning can create loop-structures and crosses.

% In the context of optimizing scaled-up device fabrication, recently developed shadow evaporation techniques of superconductors on semiconductors have shown to significantly increase device stability in superconducting quantum devices.\cite{krizek2017growth, gazibegovic2017epitaxy, carrad2020shadow, khan2020highly} Similar techniques can be integrated into the NC network platform by growing nanowires at selected positions from the networks in order to create locally specified superconductor junctions. 

% This geometry can be produced by merging three nanowires and deterministically placing shadow-nanowires such that a superconducting island is formed after deposition.

\section*{Conclusion and Outlook}

We combine two well-established techniques, growth of nanowires and microtromy, to make a scalable platform for production of electronic quantum devices. The platform is generic as it is not limited to VLS grown nanowires, but can be extended to other growth or etching-based techniques, many of which are more cost-efficient, allowing the platform to use a broad range of materials. 

% The presented NC platform is an innovative approach to producing quantum devices on single NCs and NC networks in a scalable way. 

% It combines the principles of microtomy with VLS growth of nanowires which are both well-established techniques. 

% We show here how an off-the-shelf resin acts as an extension of a target substrate allows us to use standard EBL procedures to build well-behaved FETs and electronic quantum devices. 

% The unique approach of locking nanowires in a hardened resin before cutting and transferring individual lammellas with arrays of NCs allows for free positioning of NCs while keeping track of individual units. The lamellas can be transferred to any type of substrate without the need for any target substrate preparations. 

During device fabrication, selected lamellas can be used as samples for quality control by e.g. AFM, TEM and SEM. We expect this to become a desirable feature when transitioning technologies from prototype to high through-put manufacturing. 
Based on the VLS mechanism we show how this platform can be extended to NC networks of complex shapes and be merged with semiconductor/superconductor technologies to build complex superconducting quantum devices. We expect the NC-transfer platform to be ideal for realization of large arrays of quantum dots with specifically tailored tunneling barriers, which is an interesting new route for quantum information technology and fundamental studies, such as realization of lattice-based quantum simulators and wireless single-electron logic controlled by AC-fields.\cite{hubbard1963electron, korotkov1995wireless} 

% Future work will investigate devices based on core-shell NCs, (dis)connected NC arrays and different superconducting shells, as well as alignment optimization between the nanowire high symmetry crystal axes and the microtome knife cut plane to improve crystal cleaving and NC morphologies. In addition, different avenues for improving transport signatures should be explored such as atomic layer deposition of high-$\kappa$ dielectric materials prior to microtoming, removal/change of the resin and in-situ deposited contacts.

%At our current state of research 18 lamellas can be sectioned in a row and transferred in an ensemble to cover large areas in a single transfer as further discussed in Supporting Information, Section 8, and further investigated in Ref.\cite{watson2014nanoskiving}

Routes not investigated here involve growth of radial heterostructure p-n NCs embedded in a transparent resin for production of optoelectronic devices. Additionally, stacking of multiple lamellas containing NCs with deliberately tuned p-n junctions could be used for high efficiency tandem solar cells. This principle could be further extended for extreme down-scaling inter-pixel pitch between RGB sub-units for ultra high resolution monitors. 

In conclusion, this platform provides a new approach to nanoscaled device engineering with prospects for scalable fabrication and opens up new ways to integrate nanomaterials into emerging technologies within the fields of optics, optoelectronics, electronics and quantum devices. 

\section*{Experimental Section}

\textbf{Nanowire growth.} A molecular beam epitaxy (MBE) system is used to grow Au-seeded wurtzite InAs nanowires along the [0001]B direction on InAs (111)B substrates using the vapour-liquid-solid mechanism. Arrays of Au catalyst particles are placed via standard EBL with particle radius $r_{\mathrm{Au}} =$ 20-120 nm and height $h_{\mathrm{Au}} =$ 10-50 nm. After substrate annealing at As overpressure at $T = $ 500 C$^{\circ}$ for 5 min, predominantly vertical nanowire growth is initiated at growth temperatures ranging from $T_{\mathrm{growth}} =$ 445-450 C$^{\circ}$. Axial nanowire growth is carried out for a duration of 10-120 min before a short break (5 min) is introduced and the As$_{\mathrm{4}}$/As$_{\mathrm{2}}$ ratio is increased.

After vertical nanowire growth is concluded, we switch to growth conditions favoring radial growth. First, we introduce a short pause (5 min) before the temperature of the growth substrate is lowered steadily to $T_{\mathrm{growth}}$ = 350 C$^{\circ}$ over about 15 min. After an additional pause (5 min) the nanowires are radially overgrown for 1-20 min depending on the desired amount of overgrowth and distance between Au particles. An exhaustive description can be found in recently published work.\cite{kanne2021double} \\

\textbf{Microtomy.} InAs growth substrate are cleaved into smaller pieces containing the as-grown free standing nanowire arrays. The wafer pieces are placed into silicone molds with the growth substrate facing the bottom, such that the nanowires are orientated towards the mold opening. Next, the embedding solution is prepared using the SPURR Low-Viscosity resin applying the 'hard' epoxy recipe. Ten samples are created from mixing 4.10 g ERL 4221 (cycloaliphatic epoxide resin), 0.95 g diglycidyl ether of polypropylene glycol, 5.90 g nonenyl succinic anhydride and 0.10 g dimethylaminoethanol. 
The mold is then filled with the solution before it is baked for eight hours at 70$^{\circ}$C to harden the resin. Now the growth substrate is removed from the sample using a razor blade leaving the nanowires embedded, before the block-face dimension is trimmed to accommodate our device blanks ($\sim$250 x 250 $\mu$m). The trimmed samples are mounted in a Leica Ultracut UCT UltraMicrotome and sectioned with a DiATOME Ultra 45$^{\circ}$ diamond knife at a clearance angle of 6$^{\circ}$. After sectioning the lamellas are floating on top of the fluid (ultra-pure water) contained by the knife boat. The desired substrate is partially submersed, the lamellas are guided to the desired location and the substrate withdrawn now containing the deposited lamellas. In our case lamellas were transferred to Si$^{++}$/SiO$_{\mathrm{2}}$ substrates, glass slides and TEM grids. \\

\textbf{Microscopy.} TEM characterization of the NCs in the lamellas was performed using a 200 kV Philips-FEI CM20 large tilt TEM. The plane waves are aligned parallel to the [0001] zone axis of the NCs. High-resolution TEM images were obtained using a Jeol 300F microscope. SEM characterization of the substrates with as-grown nanowires were carried out with a Jeol 7800F SEM using acceleration voltages in the range of $V_{\mathrm{acc}}$ = 1-20 kV. A Raith Eline system was used to obtain SEM images of the deposited lamellas containing NCs and finished devices using acceleration voltage, $V_{\mathrm{acc}}$ = 10 kV. The cut-face quality of the NCs post-sectioning was inspected by AFM using a Bruker Dimension Icon PT AFM in PeakForce Tapping mode.\\

\textbf{Device fabrication and measurement.} All devices are fabricated on highly doped Si$^{++}$ substrates covered by 200 nm of thermal oxide. Metallic leads to the NCs were fabricated by electron beam lithography. RF ion (Ar+) milling was performed in the metal deposition chamber prior to e-beam metal deposition of Ti and Au (5/$\sim$200 nm) to create transparent ohmic contacts to the NCs. Standard low-frequency ($<$200 Hz) lock-in measurements where performed to measure differential conductance ($V_{\mathrm{exc}}$ $\sim$ 5-20 $\mu$V). No data correction was performed on the electrical data.

\begin{acknowledgements}

This work was funded by the Danish National Research Foundation (J.E.S., K.G.-R. and J.N.), European Union’s Horizon 2020 research and innovation programme under grant agreement FETOpen grant no. 828948 (AndQC) (T.K., and J.N.) and QuantERA project no. 127900 (SuperTOP) (K.G.-R. and J.N.), Villum Foundation project no. 25310 (K.G.-R.), Innovation Fund Denmark’s Quantum Innovation Center Qubiz (J.N.), University of Copenhagen (T.K.), the Novo Nordisk Foundation project SolidQ (J.N.) and the Carlsberg Foundation (J.N.). We gracefully thank Mikelis Marnauza, Dags Olsteins, Claus B. S\o rensen, Karolis Parfenuikas and Martin Bjergfelt for helpful discussions and technical assistance.

\end{acknowledgements}

%\section*{Author contributions}

\section*{Competing interests} 

The authors declare no competing interests.

\section*{Supporting Information}
Supporting Information is available from the Wiley Online Library or from the author. All raw data and curated data sets are provided online at \url{https://erda.ku.dk/archives/335ba013276d360019209d7b80fba323/published-archive.html}.

\bibliography{ref}
\bibliographystyle{advanmatstyle}

%\newpage

%\section*{Table of Content}

% \begin{figure*}[htb!]
% \includegraphics[scale=1]{figures/Table of content_2.pdf}
% \caption{\textbf{Table of content.} An innovative platform combining microtomy with newly developed nanowire network growth techniques is presented. It enables large-scale transfer and deterministic placement of high quality nanocrystals and networks hereof to almost any substrate. Individual nanocrystals exhibit excellent field effect mobilities and quantum transport phenomena, showing the potential of the platform for large-scale architectures of quantum devices.}
% \end{figure*}

%An innovative platform combining microtomy with newly developed nanowire network growth techniques is presented. It enables large-scale transfer and deterministic placement of high quality nanocrystals and networks hereof to almost any substrate. Individual nanocrystals exhibit excellent field effect mobilities and quantum transport phenomena, showing the potential of the platform for large-scale architectures of quantum devices.

% Limits for ToC: 50-60 words + 1 figure describing the main results
% Current number of words: 55
\end{document}